\documentclass[prb,english,reprint]{revtex4-2}
\usepackage[T1]{fontenc}
\usepackage[latin9]{inputenc}
\usepackage{bm}
\usepackage{hyperref}
\usepackage{amsmath}
\usepackage{amssymb}
\usepackage{babel}
\usepackage{ulem}

\newcommand{\dd}[1]{\mathrm{d}#1\,}
\newcommand{\DD}[1]{\mathrm{D}[#1]\,}

\DeclareMathOperator{\Tr}{Tr}
\DeclareMathOperator{\tr}{tr}

\renewcommand{\vec}[1]{\bm{#1}}

\usepackage{xcolor}

\definecolor{PV-color}{rgb}{0.97,0.57,0.11}

\definecolor{SB-color}{rgb}{0.57,0.97,0.11}

\definecolor{IT-color}{rgb}{0.57,0.11,0.97}

\begin{document}

\title{Magnetoelectric effects in superconductors due to spin-orbit scattering: a non-linear $\sigma$-model description}


\author{P. Virtanen}
\affiliation{Department of Physics and Nanoscience Center, University of Jyv\"askyl\"a, P.O. Box 35 (YFL), FI-40014 University of Jyv\"askyl\"a, Finland}

\author{F. S. Bergeret}
\affiliation{Centro de F\'isica de Materiales (CFM-MPC) Centro Mixto CSIC-UPV/EHU, E-20018 Donostia-San Sebasti\'an,  Spain}
\affiliation{Donostia International Physics Center (DIPC), 20018 Donostia--San Sebasti\'an, Spain}

\author{I. V. Tokatly}
\affiliation{Nano-Bio Spectroscopy Group, Departamento de Pol\'imeros y Materiales Avanzados: F\'isica, Qu\'imica y Tecnolog\'ia, Universidad del Pa\'is Vasco (UPV/EHU), 20018 Donostia-San Sebasti\'{a}n, Spain} 
\affiliation{IKERBASQUE, Basque Foundation for Science, 48011 Bilbao, Spain}
\affiliation{Donostia International Physics Center (DIPC), 20018 Donostia-San Sebasti\'{a}n, Spain}
\affiliation{ITMO University, Department of Physics and Engineering, Saint-Petersburg, Russia}

\begin{abstract}
    We suggest a generalization of nonlinear $\sigma$-model
    for diffusive superconducting systems to account for magnetoelectric
    effects due to spin-orbit scattering.
    In the leading orders of
    spin-orbit strength and gradient expansion it includes two additional terms responsible for the spin-Hall effect and the spin-current swapping. First, assuming a delta-correlated disorder we derive the new terms from the Keldysh path integral representation of the generating functional. Then we argue phenomenologically that they exhaust all invariants allowed in the effective action to the leading order in the spin-orbit coupling (SOC). Finally, the results are confirmed by a direct derivation of the saddle-point (Usadel) equation from the quantum kinetic equations in the presence of randomly distributed impurities with SOC. At this point we correct a recent derivation of the Usadel equation that includes magneto-electric effects and does not resort to the Born approximation.
\end{abstract}

\maketitle

\section{Introduction}

Spin-orbit coupling (SOC) in solids generates a variety of well-known
effects, \cite{zutic2004,nagaosa2010,sinova2015} such as the spin and
the anomalous Hall effects, where magnetic and electric degrees of
freedom couple to each other.  Two common origins of the effects are
often considered, the intrinsic SOC due to properties
of the pure lattice, and the extrinsic SOC due to impurities.  
In superconducting materials, how the
different spin-orbital effects manifest,
and are conveniently theoretically described, is still partially not resolved.

The magnetoelectric effects associated with SOC due to extrinsic
impurity scattering have been extensively discussed in the normal
state \cite{zutic2004,nagaosa2010,sinova2015}, but in the
superconducting state have received somewhat less attention
\cite{bergeret2016,espedal2017,huang2018} compared to the intrinsic effects \cite{edelstein1995}.  In contrast, the effect of the spin-orbit scattering on spin relaxation in superconductors is well-known. \cite{abrikosov1962} Theoretically, many effects concerning diffusive electron transport in systems with impurities can be described with nonlinear $\sigma$-models, \cite{wegner1979,efetov1980,efetov1983,belitz1994} which remain convenient also when superconductivity is included \cite{finkelshtein1987,belitz1994,feigelman2000,kamenev2010}. Spin-orbit relaxation within $\sigma$-models was described early. \cite{efetov1980,efetov1983} However, it is often considered to be generated by a different scattering potential from the normal scattering. This results to an omission of magnetoelectric effects, which to our knowledge were not discussed from this viewpoint.

In this work, we consider spin-orbit scattering originating from the
same potential as the normal scattering, and from this assumption
obtain additional terms in the $\sigma$-model, within the simplest
expansion in the spin-orbit strength. The result consists of two
contributions, corresponding to spin swapping \cite{lifshits2009} and
the spin-Hall effect.  The saddle-point equation is  similar to the  Usadel equation \cite{usadel1970} derived in a previous
work \cite{bergeret2016,huang2018}. There are, however,  certain differences, which we
recognize to be due to technical issues in the previous calculations. We explicitly resolve those issues by re-deriving our
result also from the earlier kinetic equation approach.

The manuscript is structured as follows.  In Sec.~\ref{sec:magnetoel}
we derive the Keldysh non-linear  $\sigma$-model including  magnetoelectric effects, and discuss its behavior at the saddle point.  In Sec.~\ref{sec:kineq}
we present an alternative derivation of the saddle-point equation, following earlier
kinetic equation approaches. Section~\ref{sec:summary} concludes the
discussion.

\section{Magnetoelectric effects}
\label{sec:magnetoel}

We consider superconductors with spin-orbit impurity scattering, described by an action with electron fields $\psi_\alpha$ on the Keldysh contour $C$ \cite{feigelman2000,kamenev2010},
\begin{align}
    S
    &=
    S_0
    -
    \int\dd{^3r}
    U(\vec{r}) 
    \int_C\dd{t}
    \bar{\psi}_\alpha \hat{V}_{\alpha\beta} \psi_\beta
    \,,
    \\
    \label{eq:Vsoc}
    \hat{V}_{\alpha\beta}
    &=
    \delta_{\alpha\beta}
    +
    i\lambda^2
    \epsilon_{ijk}
    \sigma^{\alpha\beta}_i
    \overset{\leftarrow}{\partial}_{r_j}
    \overset{\rightarrow}{\partial}_{r_k}
    \,.
\end{align}
Here $S_0$ is the action without impurities,
$U(\vec{r})$ the disorder potential, and $\lambda$ describes the spin-orbit coupling (SOC) strength. We have here integrated by parts to move
the derivative on $U$ in the SOC term $\vec{\sigma}\cdot\nabla U\times\vec{p}$ to act on the field to the left.
Summation over spin $\alpha,\beta=\pm$ 
and dimension $i,j,k=x,y,z$ indices is implied; $\epsilon_{ijk}$
is the antisymmetric tensor and $\sigma_i$ are the spin matrices.
The action $S_0$ also contains any source fields.

We now derive a Keldysh $\sigma$-model description of the diffusive transport in this system, including additional terms describing magnetoelectric effects due to the spin-orbit scattering.
We first average over the disorder, assuming it is a Gaussian random field
with $\langle U(\vec{r})U(\vec{r}')\rangle=\frac{1}{\pi\nu\tau}\delta(\vec{r}-\vec{r}')$ where $\tau=\tau(\vec{r})$ is a scattering time (possibly spatially varying) and $\nu$ the Fermi-level density of states. Gaussian integration yields,
\begin{align}
    S
    &\mapsto
    -i\ln
    \int\DD{U} e^{-\pi\nu\tau\int\dd{^3r}U^2(\vec{r})}
    e^{i S}
    \notag
    \\
    \notag
    &=
    S_0
    +
    \int\dd{^3r}
    \frac{i}{4\pi\nu\tau}
    \Bigl(
    \int_C\dd{t}\bar{\psi}_\alpha\hat{V}_{\alpha\beta}\psi_\beta
    \Bigr)^2
    \\
    &=
    S_0
    +
    S_{\rm dis}
    +
    S_1
    +
    S_2
    \,.
\end{align}
The first disorder term $S_{\rm dis}\propto\lambda^0$ contains the quartic disorder interaction \cite{efetov1983} independent of the spin-orbit coupling. The averaging also produces additional terms related to spin-orbit scattering, $S_1\propto\lambda^2$ and $S_2\propto\lambda^4$. The $\lambda^4$ terms lead to spin relaxation and have been previously discussed in the context of $\sigma$-models. \cite{efetov1980,efetov1983}  However, the $\lambda^2$ part, responsible for  magnetoelectric effects, is often ignored. This part corresponds to diagrams with connected normal and spin-orbit scattering vertices, and obtaining them requires considering both on the same footing in the disorder average.

Decoupling the term $S_{\rm dis}$ in a previously described way \cite{efetov1983,feigelman2000}
with the local matrix field $Q_{\alpha\beta}(\vec{r};t,t')$,
leads to an action with the residual SOC interaction terms,
\begin{align}
    S
    &=
    \frac{i\pi\nu}{8\tau}\Tr Q^2
    +
    \frac{1}{2}
    \Tr \bar{\Psi}^T G^{-1}\Psi
    +
    S_1
    +
    S_2
    \,,
\end{align}
where $G^{-1}=G_0^{-1} + \frac{i}{2\tau}Q$, and $G_0$ is the Green function corresponding to $S_0$, now assumed noninteracting, decoupled with e.g. the superconducting order parameter $\Delta$ field \cite{feigelman2000,kamenev2010}.
These matrices are here defined in the Nambu-spin basis corresponding to the outer product of
$\Psi=(\psi_\uparrow,\psi_\downarrow,\bar{\psi}_\downarrow,-\bar{\psi}_\uparrow)$ and $\bar{\Psi}=-i\sigma_y\tau_x\Psi$, where $\tau_j$ indicate Pauli matrices in the Nambu space. Moreover, the retarded--advanced--Keldysh block structure is also introduced \cite{feigelman2000,kamenev2010}. Above, $\Tr$ includes integration over time and position, in addition to the matrix trace $\tr$.

We are here mainly interested in the magnetoelectric effects, for which it is sufficient to consider SOC perturbatively in the leading order in $\lambda^2$. Integrating out fermions to this order leads to
\begin{align}
    \label{eq:S}
    S
    &\mapsto
    \frac{i\pi\nu}{8\tau}\Tr Q^2
    -
    \frac{i}{2}
    \Tr\ln G^{-1}
    +
    \frac{i}{4}
    \Tr[\Sigma_1 G]
    \,,
    \\
    \Sigma_1(\vec{r},\vec{r}')
    &=
    \frac{\lambda^{2}\epsilon_{ijk}}{2\pi i \nu}
    \Bigl[
    \tau(\vec{r}')^{-1}
    \partial_{r_{i}}\delta(\vec{r}-\vec{r}')
    \sigma_{k}
    \partial_{r_{j}} G(\vec{r},\vec{r}')
    \notag
    \\&\qquad
    \label{eq:Sigma1-def}
    -
    \tau(\vec{r})^{-1}
    \partial_{r'_{i}}\delta(\vec{r}-\vec{r}')
    \partial_{r'_{j}}
    G(\vec{r},\vec{r}')\sigma_{k}
    \Bigr]
    \,.
\end{align}
Here, $\frac{i}{4}\Tr[\Sigma_1 G]=\langle{}S_1\rangle$ contains the lowest-order self-energy describing 
the magnetoelectric coupling in the Gaussian disorder model.
It has been previously discussed within the quasiclassical theory \cite{bergeret2016,huang2018}.

We next  rewrite the result in terms of the slowly varying $Q(\vec{r})=T(\vec{r})\Lambda{}T(\vec{r})^{-1}$ around a uniform saddle point $\Lambda$,
corresponding to $\lambda=0$ but including e.g. superconductivity \cite{feigelman2000}. The gradient terms associated with $\lambda\ne0$ are considered perturbatively on the same footing as those originating from the 
gradient expansion of the $\Tr\ln$ term. This  expansion is well-known and gives \cite{efetov1980,feigelman2000,kamenev2010} 
\begin{align}\label{S_0}
    S_0' = \frac{i\pi\nu}{8}
    \Tr[D(\nabla Q)^2 + 4 i \Omega Q]
    \,,
\end{align}
where $D=v_F\ell/3$ is the diffusion constant,
$\ell=v_F\tau$ the mean free path,
and $\Omega=\epsilon\tau_3 + \Delta$ contains the 
time derivative $\epsilon=i\partial_t\delta(t-t')$, and the superconducting order parameter matrix and potentially also other local self-energies in $G_0^{-1}$.
For simplicity, spherically symmetric dispersion is assumed
with Fermi velocity $v_F$. The leading contribution proportional to $\lambda^4$ describes the spin relaxation and reads explicitly as follows \cite{efetov1980,efetov1983}
\begin{equation}\label{S_2}
    S'_{2}=\frac{-i\pi\nu}{8}\Tr[\frac{1}{4\tau_{so}}Q\sigma_kQ\sigma_k]
\end{equation}
where $1/\tau_{so}=8\lambda^4p_F^4/9\tau$ is the (Elliot-Yafet) spin relaxation rate (see e.~g. \cite{bergeret2016}).

To evaluate
the magnetoelectric $\lambda^2$ term, expansion of $G$ is needed. It is conveniently obtained in the Wigner representation,
$G(\vec{r}_1,\vec{r}_2)=\sum_{\vec{p}}e^{i\vec{p}\cdot(\vec{r}_1-\vec{r}_2)}G_{\vec{p}}(\frac{\vec{r}_1+\vec{r}_2}{2})$, where
\begin{align}
    \label{eq:Sigma1-Wigner}
    S_1'&=
    \frac{i}{4}\Tr[\Sigma_1 G]
    =
    \int\dd{^3r}
    \frac{i\pi\nu\lambda^{2}\epsilon_{ijk}}{4\tau}
    \tr
    \Bigl[
    i
    \sigma_k
    \langle{p_i G}\rangle
    \langle{p_j G}\rangle
    \notag
    \\&\;
    -
    \frac{1}{2}
    \sigma_k 
    \{
    \langle{p_iG}\rangle,
    \partial_{j} \langle{G}\rangle
    \}
    +
    \frac{i}{4}\sigma_k\partial_{i}\langle{G}\rangle\partial_{j}\langle{G}\rangle
    \Bigr]
    \,,
\end{align}
and $\langle f\rangle = \frac{i}{\pi\nu}\sum_{\vec{p}}f_{\vec{p}}$.
We denote here and below $\partial_j = \partial_{r_j}$ acting on Wigner transformed functions. The Green function can be found by solving the Dyson equation:
\begin{align}
    (G_0^{-1} + \frac{iQ}{2\tau})G_{\vec{p}}
    -
    \frac{i}{2}\nabla_p G_0^{-1}  \cdot \nabla_r G_{\vec{p}}
    -
    \frac{\nabla_r Q}{4\tau} \cdot \nabla_p G_{\vec{p}}
    =
    1
    \,,
\end{align}
here expanded to first order in gradients.
Iterating the equation once leads to
\begin{align}
    G_{\vec{p}}
    &\simeq
    \mathcal{G}
    -
    \frac{i}{2}
    \Bigl(
    \mathcal{G} \vec{v}\cdot \nabla_r \mathcal{G}
    -
    \nabla_r \mathcal{G} \cdot \vec{v} \mathcal{G}
    \Bigr)
    \,,
    \\
    \mathcal{G}(\vec{r},\vec{p})
    &=
    \frac{1}{G_0^{-1}(\vec{p}) + \frac{i}{2\tau}Q(\vec{r})}
    \,,
    \;
    \vec{v} = -\nabla_p G_0^{-1}(\vec{p})
    \,.
\end{align}
The momentum sums can then be evaluated up to accuracy $1/(p_F\ell)$:
\begin{align}
    \langle{G\rangle}
    &\simeq
    Q
    \,,
    &
    \langle{p_i G\rangle}
    &\simeq
    -
    \frac{p_F\ell}{3}Q\partial_{i} Q
    \,.
\end{align}
After substitution of this result  into Eq.~\eqref{eq:Sigma1-Wigner} we  find the gradient expansion of the SOC term in leading order,
\begin{align}
    S_1'
    &\simeq
    \frac{i\pi\nu}{8}
    \int\dd{^3r}
    \Bigl(
    -iD
    \varkappa
    \epsilon_{ijk}
    \tr[
    \sigma_k
    \partial_{i} Q \partial_{j} Q
    ]
    \notag
    \\&\qquad
   \label{eq:soc-terms}
    +
    D
    \theta
    \epsilon_{ijk}
    \tr[
    \sigma_k 
    Q\partial_{i} Q \partial_{j} Q
    ]
    \Bigr)
    \,,
\end{align}
where we identify 
\begin{align}
    \label{eq:kappatheta}
    \varkappa &= \frac{2p_F^2\lambda^2}{3}
    \,,
    &
    \theta &= \frac{2p_F^2\lambda^2}{p_F\ell}
    \,,
\end{align}
as the spin-swapping and the spin-Hall (side-jump) coefficients \cite{lifshits2009} respectively.
Their values agree with Born approximation results for the scattering.
The Gaussian disorder assumption precludes obtaining the skew-scattering contribution, but it would only adjust the values of the coefficients, as we discuss below.

The forms of the terms in Eq.~\eqref{eq:soc-terms}
can also be argued phenomenologically. Firstly, since $Q^2=1$ and $\{Q,\partial_iQ\}=0$, matrix functions $F(Q,\partial{}Q)$ of second order in derivatives can be expressed as linear combinations of $\partial_i Q \partial_jQ$ and $Q \partial_i Q \partial_jQ$. Secondly, terms of the first order in spin-orbit scattering are also expected to contain traces with one Pauli matrix $\sigma_k$. Finally, the invariance under rotations requires that the coefficient tensors  are isotropic,
and must be proportional to $\epsilon_{ijk}$. Therefore we are left with only two scalar invariants allowed in the effective action, $\Tr[\epsilon_{ijk}\sigma_k\partial_i Q\partial_j Q]$ and $\Tr[\epsilon_{ijk}\sigma_k Q\partial_i Q\partial_j Q]$, which are the forms we have obtained microscopically in Eq.~\eqref{eq:soc-terms}. Similarly, one can argue that the spin relaxation term of Eq.~\eqref{S_2} is the lowest in gradients (0th order) spin-dependent contribution allowed by the time-reversal and rotation invariance.

By combining Eqs.~\eqref{S_0}, \eqref{S_2}, and \eqref{eq:soc-terms} we obtain the final effective action of the generalized nonlinear $\sigma$-model,
\begin{align}\nonumber
    S_{\rm eff} &=\frac{i\pi\nu}{8}\Tr\Big[ D(\nabla Q)^2 + 4i\Omega Q
    - \frac{1}{4\tau_{so}}Q\sigma_kQ\sigma_k \\
    &- iD\varkappa\epsilon_{ijk}\sigma_k\partial_i Q\partial_j Q
    + D\theta\epsilon_{ijk}\sigma_k Q\partial_i Q\partial_j Q\Big].
    \label{S-eff}
\end{align}
This action is the main result of the present work. It takes into account the main physical effects of extrinsic SOC -- the spin relaxation, spin Hall effect, and spin swapping. Importantly, our phenomenological arguments show that only the values of the coefficients may depend on a specific model of disorder, while the form of the action is universal, provided the SOC remains sufficiently weak. In the next section we will confirm this at the level of the saddle-point equation, by deriving it directly from the quantum kinetic equation (Kadanoff-Baym) equation and going beyond the Born approximation (equivalent to a delta-correlated disorder potential in the path integral).

In \eqref{S-eff}  we can recognize that when $D\varkappa$ is spatially
constant, the spin-swapping term in Eq.~\eqref{eq:soc-terms} is a total derivative. Hence, only $\nabla_r(D\varkappa)$ will appear in the saddle-point equations for $Q$, and its effect on spin accumulation concentrates on e.g. surfaces where the value of $D\varkappa$ varies. Note also that the spin Hall "$\theta$-term" we find above is not a total derivative. Without a spin dependence, its counterpart would be $\Tr{}[\epsilon_{ijk}b_kQ\partial_iQ\partial_jQ]$ which can exist if the system possess an axial vector $\vec{b}$. This is a well-known topological term in 2D \cite{levine1983} describing the quantum Hall effect, whereas the spinless counterpart of the "swapping term", $\Tr{}[\epsilon_{ijk}b_k\partial_iQ\partial_jQ]$, does not appear due to rotation invariance \cite{Altland-book}.

Let us now include [U(1) and/or SU(2)] vector potential source fields $a_j$.
In the leading order in the gradient and $1/p_F$ expansions in Eq.~\eqref{eq:soc-terms},
they can be added via the covariant replacement $\partial_jQ\mapsto{}\partial_jQ - i[a_j,Q]$.
The part of $S'_1$ linear in $a_j$ reads
\begin{align}
    \label{eq:source-a-soc}
    \delta S_1'
    &=
    -\frac{\pi\nu}{8}\int\dd{^3r}
    D\epsilon_{ijk}
    \tr a_i
    \Bigl[
    i\varkappa[Q\partial_jQ,\sigma_k]Q
    \notag
    \\&
    +
    \theta\{\partial_j Q,\sigma_k\}Q,
    Q\Bigr]
    \,,
\end{align}
and provides the contributions to the
(spin) current from the SOC. Therefore the total "matrix current" takes the following form
\begin{align}
    \label{J-commutator}
    \mathcal{J}_i &= -D\Bigl(Q\partial_iQ 
    \\\notag&\qquad
    - \frac{\epsilon_{ijk}}{4}
    \Bigl[i\varkappa[Q\partial_jQ,\sigma_k]Q
    + \theta\{\partial_j Q,\sigma_k\}Q,Q\Bigr]\Bigr),
\end{align}
where the first term is the usual current originating from the standard action of Eq.~\eqref{S_0}. It is instructive to rewrite Eq.~\eqref{J-commutator} in a different form,
\begin{align} \nonumber
    \mathcal{J}_i = -D\Big(Q\partial_iQ
    &- i\epsilon_{ijk}\frac{\varkappa}{4}
    [Q\partial_jQ,\sigma_k + Q\sigma_kQ] \\
    &-\epsilon_{ijk}\frac{\theta}{4}
    \{\partial_j Q,\sigma_k + Q\sigma_kQ\} \Big),
    \label{J-final}
\end{align}
which simplifies the comparison with the current entering the Usadel equation derived in Refs.~\onlinecite{bergeret2016,huang2018}. This representation explains the identification of $\varkappa$ and $\theta$ in Eq.~\eqref{S-eff} with the swapping coefficient and the spin Hall angle, respectively. In fact, Eq.~(4) of Ref.~\onlinecite{bergeret2016} is identical to Eq.~\eqref{J-final} up to the replacement $\sigma_k \mapsto \frac{1}{2}( \sigma_k + Q\sigma_kQ)$, and to Eq.~\eqref{J-commutator} up to a projection $\mathcal{J}_i\mapsto{}\frac{1}{2}(\mathcal{J}_i-Q\mathcal{J}_iQ)$. The origin of this difference will be discussed in detail in the next sections.

\subsection{Saddle point}

The saddle point equation for the action $S_{\rm eff}$ of Eq.~\eqref{S-eff} is derived in a usual way \cite{kamenev2010} by requiring stationarity of the action under the following variation $\delta Q=[w,Q]$, where $w$ is an arbitrary function, which ensures that the condition $Q^2=1$ is preserved. The result has a form of the Usadel equation \cite{usadel1970,kamenev2010,feigelman2000},
with additional gradient terms originating from the $\lambda^2$ part of Eq.~\eqref{eq:soc-terms}, 
\begin{align}
    \label{eq:sigma-sp-usadel}
    0&=\Bigl[Q,
    -i\Omega + \frac{1}{8\tau_{so}}\sigma_kQ\sigma_k
    +
    \frac{1}{2}
    \partial_k(D\partial_k Q) 
    \\\notag&\;
    -
    \frac{D\theta}{4}\epsilon_{ijk}(
    \partial_i Q\partial_j Q\sigma_k + \partial_i Q \sigma_k \partial_jQ
    + \sigma_k\partial_iQ\partial_jQ)
    \\\notag&\;
    + i\epsilon_{ijk}\frac{\partial_i (D\varkappa)}{4}[Q\partial_jQ,\sigma_k]Q
    + \epsilon_{ijk}\frac{\partial_i (D\theta)}{4}\{\partial_jQ,\sigma_k\}Q
    \Bigr]
    \,,
\end{align}
with $Q^2=1$. By construction the equation is of a commutator form, which makes it consistent with the normalization condition.
Note that $[Q,\frac{1}{2}\partial_k(D\partial_kQ)]=\partial_k(DQ\partial_kQ)$, and that only the derivative of the spin-swapping coefficient
$D\varkappa$ enters the equation due to its total derivative form in Eq.~\eqref{S-eff}.

The saddle point (Usadel) equation \eqref{eq:sigma-sp-usadel} can be rewritten in a physically more transparent form as follows,
\begin{align} \label{Usadel-saddle}
&\left[-i\Omega,Q\right] + \partial_{k}{\cal J}_k 
={\cal T}- \frac{1}{8\,\tau_{\mathrm{so}} } \left[\sigma_k Q  \sigma_k , Q  \right]
\,,
\end{align}
where ${\cal J}_k$ is the matrix current of Eq.~\eqref{J-final} [or, equivalently Eq.~\eqref{J-commutator}], and ${\cal T}$ a SOC correction to an effective torque originating from the spin Hall and the spin swapping effects \cite{bergeret2016,huang2018}
\begin{align} \label{torque-saddle}
 {\cal T} = \frac{D}{4}\epsilon_{ijk} \bigg[
 \theta\big[\sigma_k,Q \partial_{i}Q \partial_{j}Q \big]
+ i\varkappa\big[\partial_{i}Q\partial_{j}Q,\sigma_k] \bigg].
\end{align}

The saddle point equation \eqref{Usadel-saddle} is identical to the Usadel equations derived in Refs.~\onlinecite{bergeret2016,huang2018} up to one point -- an effective renormalization of the spin matrices $\sigma_k \mapsto \frac{1}{2}( \sigma_k + Q\sigma_kQ)$ in the expression for the current ${\cal J}_k$ in Eq.~\eqref{J-final}. As we will see shortly, the reason is an  inconsistency in Refs.~\onlinecite{bergeret2016,huang2018} due to neglecting normalization constraints on the  perturbative solutions of the Eilenberger equation in the diffusive limit.  A corrected calculation recovers the results above. We clarify this issue in the next section.

\section{Kinetic equation derivation}
\label{sec:kineq}

In this section we derive the Usadel equation in the presence of SOC from the quantum kinetic equation, which is a more customary way \cite{larkin1986}.  We follow here Ref.~\onlinecite{huang2018}, and restate the main points in the derivation for completeness, up to the point where differences appear.

As in the previous section, we introduce the Keldysh matrix Green functions (GF) which is $8\times8$ matrix 
\begin{gather}
{G}=\begin{pmatrix}
G^{R} & G^{K}\\
0 & G^{A}
\end{pmatrix}\,,
\end{gather}
where $G^{R,A,K}$ are the retarded, advanced and Keldysh 4$\times$4
matrices in the Nambu-spin space. ${G}$ obeys the equation
\begin{equation}
\begin{split}
\left[\tau_{3}i\partial_{t}+\frac{\nabla_{r}^{2}}{2m}+\mu+{\bf h}.\vec{\sigma}+{\Delta}-{\Sigma}\right]{G}(\vec{r},t;{\bf r'},t')
\\
=\delta({\bf r-r'})\delta(t-t')\;,\label{eq:Gorkov}
\end{split}
\end{equation}
where $\mu$ is the chemical potential, and ${\Delta}$ the
superconducting order parameter. The self-energy ${\Sigma}$
describes the impurity scattering, including the spin-orbit coupling
term. In order to obtain the quantum kinetic equation from Eq.~\eqref{eq:Gorkov}, one follows a well-known scheme: (i) Subtract from Eq. (\ref{eq:Gorkov}) its conjugate, (ii) perform the Wigner transform and then (iii) the gradient expansion \cite{larkin1986}. Following this procedure one finally obtains the kinetic equation: 
\begin{equation}
\frac{p_{k}}{m}\partial_{k}{G}_{\vec{p}}(\vec{r})+i\tau_{3}\partial_{t}{G}_{\vec{p}}(\vec{r})-i\partial_{t'}{G}_{\vec{p}}(\vec{r})\tau_{3}={\cal I}\;,\label{eq:kin_eq}
\end{equation}
where ${\cal I}$ is the collision integral, which is a functional
of the Wigner transformed ${G}_{\vec{p}}(\vec{r})$ and ${\Sigma}_{\vec{p}}(\vec{r})$ (see Eq. (18) in Ref.~\onlinecite{huang2018}).  Because the GFs are peaked at
the Fermi level it is convenient to introduce the quasiclassical GF
which is defined as ${g}(\vec{n},\vec{r})=(i/\pi)\int d\xi{G}_{\vec{p}}(\vec{r})$,
where $\vec{n}$ is a unit vector pointing in the direction of the
momentum at the Fermi surface. As in Refs.~\onlinecite{bergeret2016,huang2018},
we assume the diffusive limit and expand ${g}$ in spherical
harmonics keeping zeroth and first moments, ${g}(\vec{n},\vec{r})\approx{g}(\vec{r})+n_{k}{g}_{k}(\vec{r})$.
The two moments are determined by following equations \cite{huang2018}:
\begin{align}
\tau_{3}\partial_{t}{g}+\partial_{t'}{g}\tau_{3}+\frac{v_{F}}{3}\partial_{k}{g}_{k} &= {{\cal I}_{0}}
\,,
&
\frac{v_{F}}{3}\partial_{k}{g}&={{\cal I}_{k}}\,, \label{eq:kin-iso-aniso}
\end{align}
where 
\begin{align}  \label{eq:coll-inta}
{\mathcal{I}}_0 [{g} ,{g}_k] &=  -i \left\langle 
\left[\Sigma ,\,{\mathrm{g}} \right] \right\rangle  
 - \frac{\partial_{i}}{2} \left\langle  \big\{ \partial_{p_i} \Sigma \,, \, {\mathrm{g}}\big\}      \right\rangle  \\
 {\mathcal{I}}_{k} [{g} ,{g}_k] &= -i \langle 
 n_k \left[ \Sigma ,\,{\mathrm{g}} \right]\rangle \label{eq:coll-intb}\;.
\end{align}
Here $\langle\cdot\rangle$ indicates average over momentum direction $\vec{n}$. These collision integrals have been evaluated in Ref.~\onlinecite{huang2018} by expanding the  self-energy ${\Sigma}$  in terms of the small parameter  $\lambda^2 p_{F}^2$ up  to second order:
\begin{equation}
{\Sigma} = {\Sigma}^{(0)}+ {\Sigma}^{(1)} + {\Sigma}^{(2)},
\label{eq:Sigma_all}
\end{equation}
The zeroth order self-energy describes the usual elastic relaxation. The first and second order describe spin-charge coupling and the spin relaxation process respectively. The former was discussed in the previous section. After a lengthy, but straightforward, calculation one can show that Eqs.~\eqref{eq:kin-iso-aniso} can be written as: \cite{huang2018}
\begin{align} \label{eq:USA}
&\left[-i\Omega,  g\right] + \partial_{k}\left(\frac{v_F}{3}{g}_k+{\cal J}_k^{an}\right) 
={\cal T}- \frac{1}{8\,\tau_{\mathrm{so}} } \left[\sigma_a {g}  \sigma_a ,  {g}  \right]\\
& \frac{v_{F}}{3} \partial_k {g} + \big[ {\mathcal{A}}_k \, , \,  {g} \big] =0,\label{eq:USA1}
\end{align}
where $i{\Omega}=i\epsilon\tau_{3} + i{\Delta}$ and
\begin{align}
{\mathcal{J}}^{\mathrm{an}}_k 
&=  
\frac{D}{2} \epsilon_{akj}
\bigg[ \frac{\omega_1  \tau}{p_{F} l }\,\big\{\partial_{j}{g} ,\sigma_{a}\big\}+ i \frac{\omega_2 \tau}{p_Fl }  [\sigma_a, {g} \partial_j {g} ]\bigg],\label{eq:Jan}
\\
{\cal{T}}
&=\frac{D}{4}\epsilon_{akj} \bigg[
 \left( \frac{2}{3} \omega_2 \tau + \frac{2 \omega_1 \tau  }{p_F l } \right) \,\big[\sigma_{a},{g} \partial_{k}{g} \partial_{j}{g} \big]
\label{eq:torque}
\\
&\quad
+\left(\frac{2}{3}\omega_1\tau -\frac{2\omega_2 \tau}{p_Fl}  \right)
i\big[\partial_{k}{g}\partial_{j}{g},\sigma_{a}] \bigg]
\notag
\end{align}
are the matrix anomalous current and spin-orbit torque terms. They are defined
in terms of the spin-charge coupling rates
$\omega_1$ and $\omega_2$ which are related to the components of the single-impurity scattering matrix at the Fermi energy: $ \hat{t}_{\vec{p}\vec{p}'}=A + i(\vec{p}\times\vec{p}')\cdot \bm\sigma B/p_F^2$ via $\omega_1 = 2\pi n_{\mathrm{im}} N_F \mathrm{Re} \big[ A^* B\big]$ and  $\omega_2= 2\pi n_{\mathrm{im}} N_F \mathrm{Im}\big[A^* B\big]$. \cite{huang2018} 
Moreover, the matrix ${\cal A}_k$ in Eq.~\eqref{eq:USA1} 
is defined as
\begin{align}  
{\mathcal{A}}_k &= \frac{{g}_k}{6\tau} + \frac{\omega_1 \epsilon_{ajk}}{6} \left( \frac{1}{3} \,i \left[{g}_{j} ,\sigma_{a}\right] - \frac{1}{2p_{F}} \left\{ \partial_{j}{g} ,\sigma_{a}\right\} \right) \nonumber \\
&+ \frac{\omega_2 \epsilon_{ajk}}{12p_F}   \left(   \sigma_a i\partial_j {g}\, {g} + {g}\,i \partial_j {g} \sigma_a \right) \nonumber \\
&+ \frac{\omega_2 \epsilon_{ajk}}{18}   \left( \sigma_a {g}_j {g}    - {g}\, {g}_j \sigma_a \right).\label{eq:Akk}
\end{align}
Equations~(\ref{eq:USA}--\ref{eq:USA1}) form a closed system of equations for the zeroth, $ g$, and the first moment, ${g}_{k}$, 
of the GF. The second equation allows expressing ${g}_{k}$
in terms of $ g$, and after substitution in the first equation,
one obtains the Usadel equation. The structure of Eq.~\eqref{eq:USA1}, ensures on the one hand the
normalization condition
$ g^{2}=1$,  and on the other hand that $\partial_{k}( g\partial_{k}g)$
can be represented as a commutator $\left[..., g\right]$. In other
words, independently of the way used to derive it, the Usadel equation
must have the commutator structure 
\begin{equation}
[\Lambda, g]=0\;,\label{eq:comm_form}
\end{equation}
where $\Lambda$ is a certain matrix. 

Consequently, one can now suspect that the Usadel equation derived in Ref.~\onlinecite{huang2018} is not correct, because it does not preserve the commutator form.  The cause of this inconsistency  is  in the procedure for solving the equation system~(\ref{eq:USA}--\ref{eq:USA1}). This  can be corrected to obtain a consistent solution, as we discuss next. 

The procedure is to express ${g}_k$ in terms of $ g$ using Eq.~\eqref{eq:USA1}, and then substitute this expression into Eq.~\eqref{eq:USA}. 
We work in the leading order in small SOC and write the first moment as
\begin{equation}
 g_{k}=-l g\partial_{k} g+\delta  g_{k}\label{eq:first moment}
\end{equation}
 where $\delta  g_{k}$ is the correction due to SOC.  From Eq.
(\ref{eq:USA1}) we get the equation for $\delta  g_k$ 
\begin{equation} \label{eq:temp1}
\bigg[  {g}\, , \,
\frac{ \delta {g}_{k}}{6\tau} - \frac{v_F \epsilon_{akj}}{12}  
\left(  \theta'  \left\{\partial_{j}{g} ,\sigma_{a}\right\} 
+i \varkappa'  \left[{g} \partial_{j}{g} ,\sigma_{a}\right] \right)
 \bigg]=0,
\end{equation}
where the parameters $\theta'$ and $\varkappa'$ are defined as follows
\begin{align}
\varkappa' &=\frac{2}{3}\omega_1 \tau  -\frac{\omega_2  \tau }{p_Fl }\,,
&
\theta' &=  \frac{2}{3}\omega_2 \tau  + \frac{\omega_1  \tau }{p_F l }\,,
\label{eq:l_Hkappa}
\end{align}
The commutator equation~\eqref{eq:temp1} has multiple solutions. 
In Ref.~\onlinecite{huang2018}, the authors choose the solution for $\delta g_{k}$
which nullifies the second term in the commutator. It is this choice that in the end leads to an equation which does not have the commutator structure,   and hence does not ensure the normalization condition.
In order to obtain the correct Usadel equation one can note that the general solution of Eq.~\eqref{eq:temp1} can be written
as
\footnote{
  To see that this is the general solution,
  one can note that $[{g},X]=0$ obtains the simple form
  $[\tau_3,\tilde{X}]=0$ in the eigenbasis of ${g}$,
  after which the statement follows.
}
\begin{align}
\delta  g_{k}=\frac{l \epsilon_{akj}}{2}  
\left(  \theta'  \left\{\partial_{j}{g} ,\sigma_{a}\right\} 
+i \varkappa'  \left[{g} \partial_{j}{g} ,\sigma_{a}\right] \right)+\left\{ \Gamma_{k}, g\right\} \label{eq:sol_gk}
\end{align}
where $\Gamma_{k}$ has to be determined by imposing, that after substitution in Eq. (\ref{eq:USA}), one obtains the Usadel equation with the form of Eq. (\ref{eq:comm_form}).

To find the value of $\Gamma_k$, let us focus on the derivative term on the l.h.s of Eq.~\eqref{eq:USA}. It is a total divergence that defines, after substitution of Eqs.~\eqref{eq:Jan}, \eqref{eq:first moment},  and \eqref{eq:sol_gk}, the total current
${\cal J}_k=\frac{v_F}{3}{g}_k+{\mathcal{J}}^{an}_k$:
\begin{align}
{\cal J}_k&=-D g\partial_k  g +\frac{D \epsilon_{akj}}{2}  
  \theta \left\{\partial_{j}{g} ,\sigma_{a}\right\} + \nonumber \\
  &+ \frac{D \epsilon_{akj}}{2}   i \varkappa \left[{g} \partial_{j}{g} ,\sigma_{a}\right]
+\frac{v_F}{3}\left\{ \Gamma_{k}, g\right\}  \label{eq:tot_current0}
  \,,
\end{align}
where $\theta \equiv\theta' +\frac{\omega_1 \tau }{p_F l}=\frac{2}{3}\omega_2 \tau  + 2\frac{\omega_1  \tau }{p_F l}$
and $\varkappa\equiv\varkappa' +\frac{\omega_2  \tau }{p_Fl }= \frac{2}{3}\omega_1 \tau  -2\frac{\omega_2  \tau }{p_Fl }$
are the spin Hall angle and spin swapping coefficient,
which reduce to Eqs.~\eqref{eq:kappatheta} in the Born approximation.

The derivative term in Eq.~\eqref{eq:USA} has now a form $\partial _k(D\theta A_k+D\varkappa B_k + (v_F/3)\{\Gamma_k, g\})$. Notice that the derivative acts also on the kinetic coefficients.  The matrix $\Gamma_k$ is then obtained by imposing that such terms $\propto\partial_k(D\kappa)$, $\partial_k(D\theta)$ have the   commutator form, Eq.~\eqref{eq:comm_form} \footnote{In general, if $ g^2=1$ and one searches for a matrix $ \Gamma$ such that $ A +\{ \Gamma, g\}$ has a commutator form, then $\Gamma=-(1/2) A g$}:
\begin{align}
 \Gamma_k&= -\frac{l \epsilon_{akj}}{4}  
  \theta  \left\{\partial_{j}{g} ,\sigma_{a}\right\}{g} - 
  \frac{l \epsilon_{akj}}{4} i\varkappa  \left[{g} \partial_{j}{g} ,\sigma_{a}\right] {g}  
  \,,
\end{align}
Thus, finally we we can write the expression for the current as
\begin{align}
{\cal J}_k&=-D g\partial_k  g +\frac{D \epsilon_{akj}}{4} \left[
  \theta \left\{\partial_{j}{g} ,\sigma_{a}\right\} g +i\varkappa  \left[{g} \partial_{j}{g} ,\sigma_{a}\right] g,  g\right] 
  \label{eq:tot_current}
\end{align}
This result is identical to Eq.~\eqref{J-commutator} and therefore the form of Eq.~\eqref{J-final} with the "renormalized" spin matrices. We can note that the resulting spin-orbit terms in Eq.~\eqref{eq:tot_current} are exactly those generated by Eq.~\eqref{eq:source-a-soc}, and  originated from the covariant derivatives in the $\sigma$-model action.

With the form of ${\mathcal{J}}_k$ now fixed, the Usadel equation reads
\begin{align} \label{eq:Usadel}
&\left[-i\Omega, g\right] + \partial_{k}{\cal J}_k 
={\cal T}- \frac{1}{8\,\tau_{\mathrm{so}} } \left[\sigma_a {g}  \sigma_a ,  {g}  \right]
\,,
\end{align}
and exactly coincides with the saddle point equation \eqref{Usadel-saddle}. From here we can already conclude that this equation can be expressed in a commutator form. It is however instructive to check this directly, and indeed using antisymmetry of $\epsilon_{ijk}$ and the normalization condition ${g}^2=1$, one straightforwardly brings the equation to the form:
\begin{align}
&0 =
\Biggl[
{g}\;,
\;
-i{\Omega} - \frac{1}{8\tau_{so}}\sigma_k g \sigma_k
+ \partial_k\frac{D}{2}\partial_k {g}
\\\notag
&+\epsilon_{akj}\frac{D\theta}{4}\left(\partial_j g \sigma_a\partial _k  g+\sigma_a\partial_j g \partial _k g +\partial_j g \partial _k g \sigma_a  \right)
\\\notag
&+ \epsilon_{akj}\frac{\partial_k (D \theta) }{4} 
 \left\{\partial_{j}{g} ,\sigma_{a}\right\} g +i\epsilon_{akj}\frac{\partial_k (D \varkappa)}{4} \left[{g} \partial_{j}{g} ,\sigma_{a}\right] {g}
\Biggr]
\end{align}
This result coincides with the commutator form Eq.~\eqref{eq:sigma-sp-usadel} of the saddle-point equation for the $\sigma$-model defined by Eq.~\eqref{S-eff}. It is worth emphasizing that in the kinetic derivation of the present section we have treated the scattering of a single impurity exactly \cite{huang2018},  far beyond the model of $\delta$-correlated disorder adopted in Sec.~II. This  only leads to changes of  the  coefficients, whereas the structure of the saddle point equation, Eq. (\ref{Usadel-saddle}), and hence of the underlying $\sigma$-model,  remains unchanged, in agreement with our symmetry-based arguments in section II.

\section{Conclusions}
\label{sec:summary}

We have derived terms originating from spin-orbit impurity scattering in the Keldysh non-linear $\sigma$-model action for superconducting systems, which are the source of  magnetoelectric effects. The saddle-point equation of the resulting action is the Usadel equation, Eq. \ref{Usadel-saddle},  which includes effects proportional to the spin swapping coefficient and the spin-Hall angle.  We have  also discussed a way to derive the Usadel equation via the  kinetic equation approach, noting corrections to previously obtained results. Our  findings provide a general approach for describing magnetoelectric effects due to extrinsic spin-orbit scattering in diffusive superconductors, both in and out of equilibrium. The approach is also amenable for considering fluctuation effects,  away from the saddle point, in system with spin-orbit coupling. 

\acknowledgments

P.V. and F.S.B. acknowledge funding from EU's Horizon 2020 research and
innovation program under Grant Agreement No. 800923 (SUPERTED). I.V.T. acknowledges support by Grupos Consolidados UPV/EHU del Gobierno Vasco (Grant No. IT1249-19). F.S.B.  acknowledges funding by the Spanish Ministerio de Ciencia, Innovacion y Universidades (MICINN) (Project FIS2017-82804-P).

\bibliography{nlsmsoc}

\end{document}